\documentclass[a4paper,amsmath,amssymb,aps,prl,twocolumn,longbibliography,english,superscriptaddress]{revtex4-1}
\usepackage{amssymb}
\usepackage{graphicx}
\usepackage{dcolumn}
\usepackage{bm}
\usepackage{amsmath}
\usepackage{subfigure}
\usepackage{float}
\usepackage{color}
\usepackage[colorlinks=true,citecolor=blue]{hyperref}

\begin{document}


\title{Many-body nonequilibrium dynamics in a self-induced Floquet system}

\author{Yuechun Jiao} 
\affiliation{State Key Laboratory of Quantum Optics and Quantum Optics Devices, Institute of Laser Spectroscopy, Shanxi University, Taiyuan 030006, China}
\affiliation{Collaborative Innovation Center of Extreme Optics, Shanxi University, Taiyuan 030006, China}

\author{Yu Zhang}
\affiliation{State Key Laboratory of Quantum Optics and Quantum Optics Devices, Institute of Laser Spectroscopy, Shanxi University, Taiyuan 030006, China}

\author{Jingxu Bai}
\affiliation{State Key Laboratory of Quantum Optics and Quantum Optics Devices, Institute of Laser Spectroscopy, Shanxi University, Taiyuan 030006, China}
\affiliation{Collaborative Innovation Center of Extreme Optics, Shanxi University, Taiyuan 030006, China}

\author{Suotang Jia}%
\affiliation{State Key Laboratory of Quantum Optics and Quantum Optics Devices, Institute of Laser Spectroscopy, Shanxi University, Taiyuan 030006, China}
\affiliation{Collaborative Innovation Center of Extreme Optics, Shanxi University, Taiyuan 030006, China}

\author{C. Stuart Adams}
\affiliation
{Joint Quantum Centre (Durham-Newcastle), Department of Physics, Durham University, Durham, DH1 3LE, United Kingdom}

\author{Zhengyang Bai}
\email{zhybai@lps.ecnu.edu.cn}
\affiliation{State Key Laboratory of Precision Spectroscopy,
East China Normal University, Shanghai 200062, China}

\author{Heng Shen}%
\email{hengshen@sxu.edu.cn}
\affiliation{State Key Laboratory of Quantum Optics and Quantum Optics Devices, Institute of Opto-Electronics, Shanxi University, Taiyuan 030006, China}
\affiliation{Collaborative Innovation Center of Extreme Optics, Shanxi University, Taiyuan 030006, China}

\author{Jianming Zhao}%
\email{zhaojm@sxu.edu.cn}
\affiliation{State Key Laboratory of Quantum Optics and Quantum Optics Devices, Institute of Laser Spectroscopy, Shanxi University, Taiyuan 030006, China}
\affiliation{Collaborative Innovation Center of Extreme Optics, Shanxi University, Taiyuan 030006, China}

\date{\today}

\begin{abstract}
Floquet systems are periodically driven systems. In this framework, the system Hamiltonian and associated spectra of interest are modified, giving rise to new quantum phases of matter and nonequilibrium dynamics without static counterparts. Here we experimentally demonstrate a self-induced Floquet system in the interacting Rydberg gas. This originates from the photoionization of thermal Rydberg gases in a static magnetic field. Importantly, by leveraging the Rydberg electromagnetically induced transparency spectrum, we probe the nonequilibrium dynamics in the bistable regime and identify the emergence of a discrete time crystalline phase. Our work fills the experimental gap in the understanding the relation of multistability and dissipative discrete time crystalline phase. In this regard, it constitutes a highly controlled platform for exploring exotic nonequilibrium physics in dissipative interacting systems.

\end{abstract}


\maketitle

\begin{figure*}[htbp]
\centering    
\includegraphics[width=\linewidth]{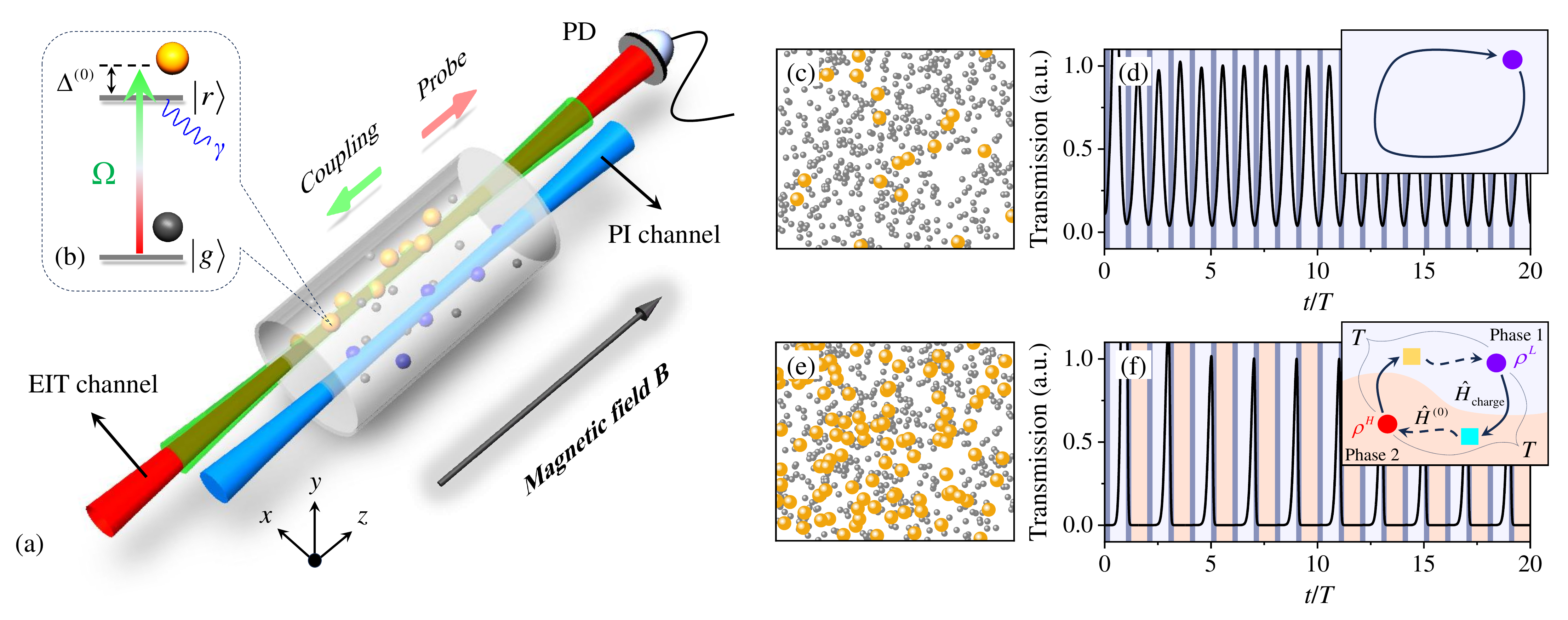}
\caption{Floquet engineering of nonequilibrium states in a driven-dissipative Rydberg gas. 
(a) Experimental setup. The probe (red) and coupling lasers (green) are counterpropagated through the Cs vapor cell to form EIT. The transmission of the probe is detected by a photodetector (PD). A PI laser beam (blue) co-propagates in parallel with the coupling laser, ionizing the Rydberg atoms in the blue channel. A homogeneous magnetic field \textit{B} is along with the probe direction. 
Gray, golden, and blue spheres represent the ground atoms, Rydberg atoms, and charged particles, respectively. (b) This setting can be treated as an effective two-level atom (ground and Rydberg states $|g\rangle$ and $|r\rangle$) with detuning $\Delta^{(0)}$, Rabi frequency $\Omega$ and decay rate $\gamma$. (c) and (d) In the non-interaction regime (with $\bar{V}=0$), the dissipation dominates the dynamics of the system and leads to a homogeneous phase (with a single stable state). As shown in (d), the electric fields driving with a period $T$ (highlighted with the light blue regime) make the system oscillate.
(e) and (f) For intense driving laser, strong atomic interactions can induce bifurcation and result in optical bistability. In the bistable regime, the phase space consists of two BOA (highlighted with gray and orange regime). 
When starting from $\rho^{L}$ (low transmission), for the one period, the charges kick the system into the  BOA of $\rho^{H}$, and subsequent systemic Hamiltonian $\hat{H}^{(0)}$  bring the system close to $\rho^{H}$ (high transmission), and vice versa. As shown in (f), in this regime, the periodic driving makes the system oscillate between high- and low- transmission with a doubled period $2T$. 
Here we set  $\bar{V}=-12\gamma$, $\Omega=0.7\gamma$, and $\Delta^{(0)}=3.5\gamma$.}\label{Fig1}
\end{figure*}

\textit{Introduction}--Understanding the complex behavior of nonequilibrium dynamics of many-body systems is a major endeavor in modern physics~\cite{eisert2015}. A paradigmatic approach to exploring nonequilibrium dynamics is to tailor the system by a periodic drive, so-called Floquet engineering. This opens a new dimension to engineer the Hamiltonian of interest~\cite{eckardt2017,weitenberg2021a,bordia2017}, and thus allows to explore the exotic phenomena inaccessible with static systems. Recent decade has witnessed great progress in uncovering Floquet phases of matter with unique space-time order or topological properties with the solid-state and cold-atom systems, such as artificial gauge fields~\cite{goldman2014,cooper2019,dalibard2011}, periodically-driven time crystals~\cite{Zhang2017,Choi2017, Kyprianidis2021, Randall2021, Kessler2021, vu2023dissipative,liu2024i} and Floquet topological insulators~\cite{oka2019,kitagawa2010,lindner2011,rudner2020}. All previous demonstrations relied on the periodic driving externally. Recently, a self-induced Floquet behavior has been theoretically investigated in metals~\cite{rudner2019}, which brings about the appearance of new phases of matter.

Rydberg atom (principal quantum number $n\gg 1$) has emerged as a simple but powerful platform for exploring novel nonequilibrium dynamics. In particular, the huge size and large electric dipole moment of highly-excited atoms leads to strong long-range van der Waals interaction and dipole-dipole interactions, fulfilling the prerequisite on complex interaction and displaying rich many-body dynamics, such as ergodicity breaking~\cite{serbyn2021quantum, bluvstein_controlling_2021,Ding2024}, self-organized criticality~\cite{Marcuzzi2016,helmrich2020signatures,Ding2020a}, optical bistability~\cite{Antiferromagnetic_Lee_2011,Carr2013, weller2016}, and collective oscillation~\cite{Wadenpfuhl2023,Wu2024}. Going beyond the external driving paradigm, in this Letter we develop a self-induced Floquet system in the interacting dissipative Rydberg gases at room temperature by exploiting a photoionization process in a static magnetic field. Building upon this periodically driven open quantum system, we observe the discrete time-crystalline (DTC) phase, which is exhibited in a ladder-type Rydberg electromagnetically induced transparency (EIT) spectroscopy. Importantly, we experimentally unveil the fact that DTC phase relies on the emergence of the bistability. 

\begin{figure}[htbp]
\centering
\includegraphics[width=\linewidth]{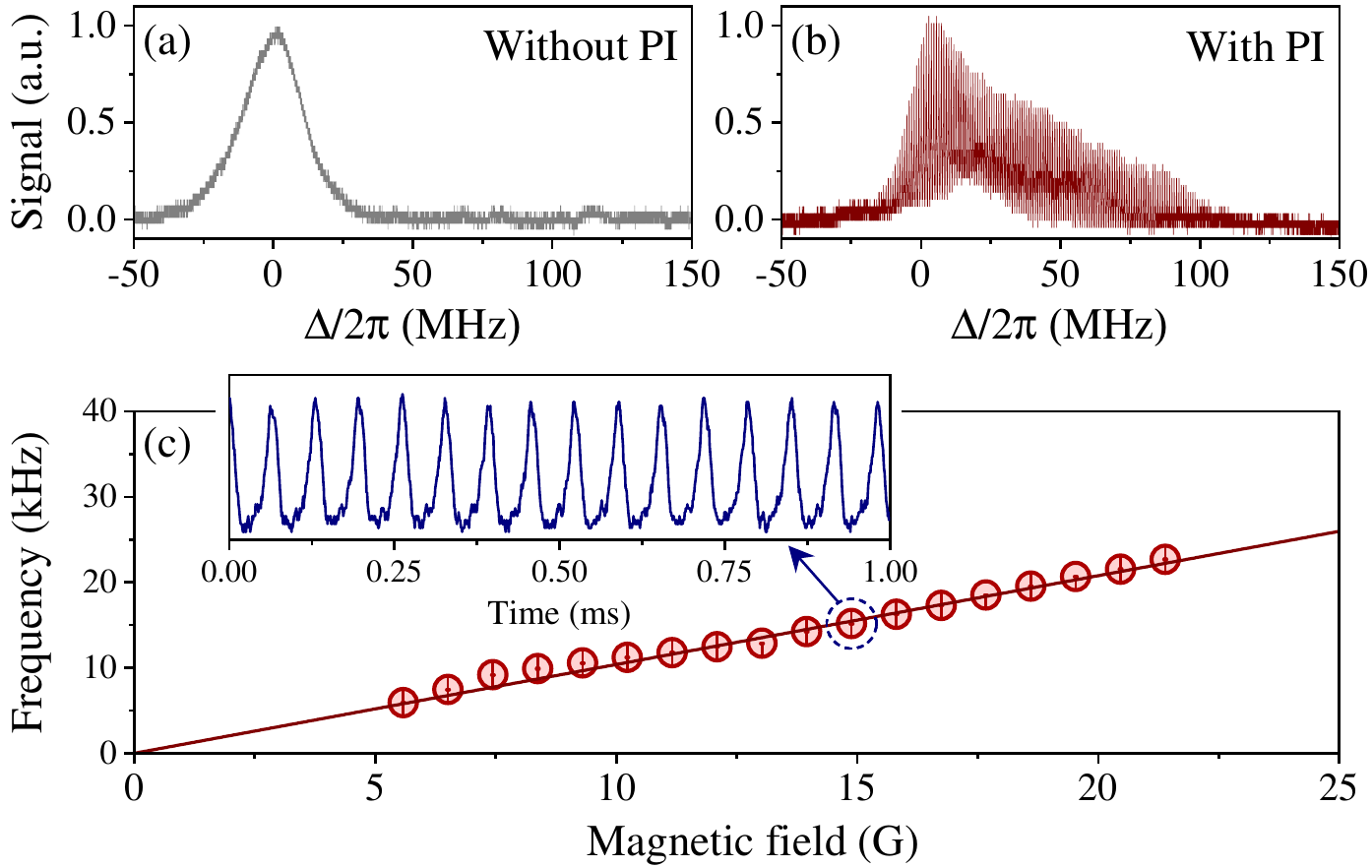}
\caption{The comparison of EIT signal without (a) and with (b) PI laser at coupling Rabi frequency of $\Omega_c/2\pi = $0.59~MHz. (c) Measured oscillation frequency as a function of the magnetic field \textit{B}. The solid line displays the linear fitting through the origin. The error bars show the standard deviation of three independent measurements. Inset presents a 1~ms time dynamics of the probe transmission 
at $\textit{B}=14.88$~G marked with a blue circle.}
\label{Fig2}
\end{figure}

We consider a cylindrical glass cell contains a vapor of caesium (Cs) atoms. As outlined briefly in Figs.~\ref{Fig1}(a) and (b),  two counter-propagated probe and coupling lasers in one optical channel excite the ground state $|g\rangle$ to Rydberg state $|r\rangle$ via an intermediate state $|e\rangle$, establishing a Rydberg EIT configuration. A spatially separated photoionization (PI) laser is applied to ionize the Rydberg atoms that diffuse from the EIT channel with a relatively long lifetime (up to 100 $\mu s$). In the presence of the magnetic field along the light propagation direction, generated charge particles are trapped by the \textit{B} field, whose number is also increased via collision with Rydberg atoms. Those charges produce significant electric fields to shift the resonance condition of the Rydberg excitation via the Stark effect. All the facts consequently induce periodically time-varying electric fields, which drive the spatially separated Rydberg EIT channel with a period of $T$. We thereby build up a self-induced Floquet system, and further use it to explore the nonequilibrium dynamics in the bistable regime as inspired by Ref. \cite{gambetta2019DiscreteTimeCrystals}. Note that although, a complete description of the dynamics would require knowledge of ion and electron motion, induced electric fields, atom-charge collisions, etc, our relatively simple description of the self-induced Floquet system is well supported by the data of oscillation frequency as a function of applied DC magnetic field [Fig. \ref{Fig2} (c)] and PI laser intensity [Fig. S4 (a)].

\textit{Model}--To investigate the Floquet-engineered system, we begin by considering an effective two-level atom. The ground state $|g\rangle$ and Rydberg states $|r\rangle$ are coupled by a laser with Rabi frequency $\Omega$ and  detuning $\Delta^{(0)}$ from resonance [see Fig.~\ref{Fig1}(b)]. The time evolution of the dissipative Rydberg system is modeled by using a time-dependent many-body Lindblad master equation $\dot{\rho}={\cal L}(\rho)$, where the generator $ {\cal L}(\cdot)=-i[\hat{H}^{(0)}+\hat{H}_{\rm charge}(t),(\cdot)]+\gamma\sum_{j}( J_j(\cdot)J_j^\dagger-\frac{1}{2}\{J_j^\dagger J_j,(\cdot)\})$ with jump operator $J_j=|g_j\rangle\langle r_j|$, and Hamiltonian $\hat{H}^{(0)}$,

\begin{eqnarray} \label{Hami}
	\hat{H}^{(0)}&&=
	\sum_{j} \left(-\Delta^{(0)}\hat{n}_j+\Omega\hat{\sigma}_j^x\right) +\sum_{j<k}{V}_{jk}\hat{n}_j\hat{n}_k,
\end{eqnarray}
where ${V}_{jk}$ represents the two-body interactions between Rydberg atoms locating at sites $\mathbf{R}_j$ and $\mathbf{R}_k$. Operator $\hat{\sigma}_j^x=|r_j\rangle\langle g_j|+|g_j\rangle\langle r_j|$ flips the atomic state and $\hat{n}_j=|r_j\rangle\langle r_j|$ is Rydberg density operator of the $j$-th atom. For the dissipative term, the loss of coherence is phenomenologically described by the decay of the Rydberg state with a rate $\gamma$. The time-dependent Hamiltonian $\hat{H}_{\rm charge}(t) = -\sum_{j}\Delta_{c}(t)\hat{n}_j$ arises from the periodic electric fields driving with $\hat{H}_{\rm charge}(t)=\hat{H}_{\rm charge}(t+nT)$. For a uniform spatial distribution of atoms in the current experiment, due to the dissipation in thermal gases, many-body correlations are weak, such that we could employ mean-field (MF) approximations to simulate the dynamics. Applying
the continuous density approximation, this yields the nonlinear MF equations of motion in the thermodynamic limit,
 \begin{subequations}\label{MF_Eq}
	\begin{eqnarray}
		&&\frac{d{\sigma}_x}{dt}=\left[\Delta(t)+\bar{V}{n_r}\right] {\sigma}_y-\frac{\gamma}{2}{\sigma}_x,\\
		&&\frac{d{\sigma}_y}{dt}=2\Omega(2{n_r}-1)-(\Delta(t)+\bar{V}n_r){\sigma}_x-\frac{\gamma}{2}{\sigma}_y,\\
		&&\frac{d{n_r}}{dt}=-\Omega {\sigma}_y-\gamma{n_r},
	\end{eqnarray}
\end{subequations}
where $\bar{V}$ is the MF energy shift, the time-dependent detuning $\Delta(t)=\Delta^{(0)}+\Delta_c(t)$, and $\sigma_{\mu}=\langle {\hat \sigma}^{\mu} \rangle \,(\mu=x, y, z)$ is the mean value of  the Pauli operator, and $n_r$ represents the Rydberg population.
In the absence of the periodic electric fields driving (e.g. $\Delta_c(t)=0$), the system behaves as optical bistability at the MF level and has been observed in the experiment~\cite{Carr2013, weller2016, Ding2020a} [see Sec. IA of Supplemental Material (SM) for details]. The many-body dynamics is governed by Hamiltonian $\hat{H}^{(0)}$. In the bistable regime, the system undergoes discontinuous phase transition when the Rydberg number density exceeds a critical value. At the critical point, the two stable states $\rho^{L(H)}$ with low-  and high-Rydberg density coexist on equal terms. 
The system can asymptotically approach one of the two stable states, which is associated with the corresponding basin of attraction (BOA). 
When applying the magnetic field, the generated periodic electric fields influence the excitation of the Rydberg atom. Consequently, they provide a periodic sequence of ``kicks'', and drive the system out of equilibrium. As shown in Fig.~\ref{Fig1}(e) and (f), we set the parameters in the bistable regime with  $\bar{V}=-12\gamma$, $\Omega=0.7\gamma$, and $\Delta^{(0)}=3.5\gamma$ (see Fig.~S2 in \textbf{SM} for MF phase diagram), 
and the system starts from the low-density stable state $\rho^{L}$ that corresponds to low transmission of the probe field (the transmission signal is proportional to Rydberg atom populations through EIT scheme; see Section IB of \textbf{SM} for details). When the electric fields interact with Rydberg atoms,
it disrupts the balance of system with an action time $t_{\rm charge}\ll t_s$ and displace it into the BOA of $\rho^{H}$ ($t_s$ is the relaxation time of the system). Subsequently, the population of charges is depleted via recombination and collisions with the walls of the cell (e.g., $\hat{H}_{\rm charge}(t)=0$), the dynamics is governed again by the Hamiltonian $\hat{H}^{(0)}$ with a time $(T-t_{\rm charge})\gg t_s$ and bring the system to its metastable regime and therefore approach to $\rho^{H}$ (corresponding to high transmission) after the first period $T$. For the second period, the action of charges can then bring the system into the BOA of $\rho^{L}$  and the $\hat{H}^{(0)}$ will carry it back to $\rho^{L}$.  
Thus, combining both Rydberg interactions and Floquet engineering by the periodic electric fields driving, the Rydberg system can 
oscillate between high- and low- transmission with a doubled period $2T$ and to the emergence of DTC phase~\cite{gambetta2019DiscreteTimeCrystals}. Outside this region (e.g., tuning system to the non-interaction situation with $\bar{V}=0$), the system only has one single stable fixed point, thus the oscillation period becomes $T$ [see  Fig.~\ref{Fig1}(c) and (d); also find detail in SM].

\textit{Experiments}--To explore the nonequilibrium dynamics mentioned above, we perform the experiments by using resonant two-photon excitation in a Cs vapor cell with a size of $\phi$ 2.5 cm $\times$ 7.5 cm. The probe (852~nm, $\Omega_p$) and coupling  (509~nm, $\Omega_c$) lasers excite the ground state $|g\rangle = |6S_{1/2}, F = 4, m_F = 4\rangle$ to Rydberg state $|r\rangle$ = $|60D_{5/2}, m_j = 5/2\rangle$ via an intermediate state $|e\rangle$ = $|6P_{3/2}, F^\prime = 5, m_F^\prime = 5\rangle$ with detuning of $\delta_p$ = 2$\pi \times$110~MHz. Both lasers are $\sigma^+$ circular polarization. The coupling laser works in the frequency-scanning mode at a typical scan rate 2$\pi\times$ 10~MHz/ms, or in the locking mode with a fixed laser detuning through a tunable offset-lock frequency technique. The $1/e^2$ beam waist of 852~nm and 509~nm are $\omega_p$ = 500~$\mu$m and $\omega_c$ = 425~$\mu$m, respectively.
The PI laser at 509 nm is 110 MHz blue detuned to the coupling laser and propagates in parallel with the coupling laser given a spatial separation of 6~mm. A homogeneous magnetic field \textit{B} is applied along with the probe direction. Experimentally, we adopt \textit{B} = 11.6~G, PI laser of 200~mW, and $\Omega_p / 2 \pi =$ 25.1~MHz in the experiments unless otherwise stated.

\begin{figure*}[htbp]
    \centering
\includegraphics[width=\linewidth]{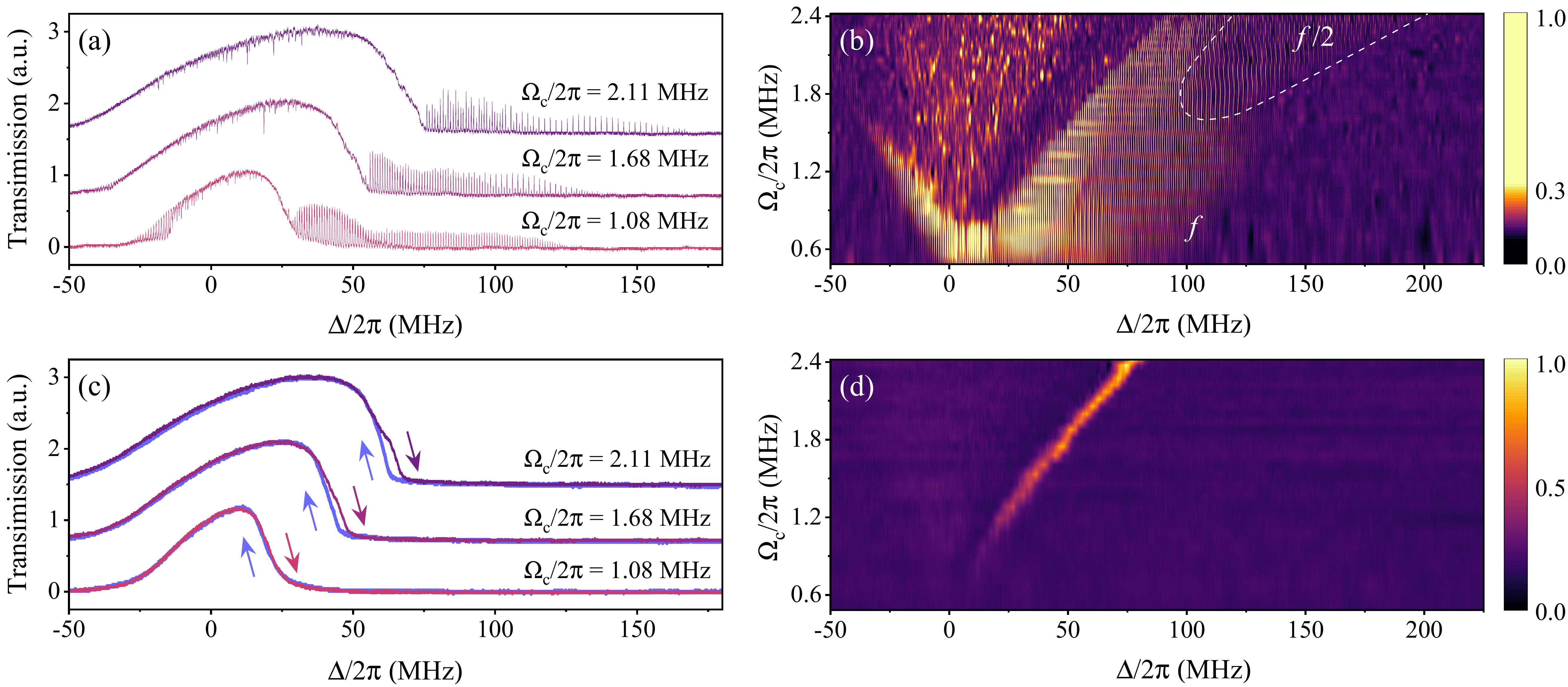}
    \caption{(a) The measurement of scanned EIT signal at the indicated Rabi frequency of $\Omega_c/2\pi = $1.08~MHz, 1.68~MHz and 2.42~MHz, respectively. (b) Measured color map of the probe transmission as a function of $\Delta$ and Rabi frequency of $\Omega_c$. (c) The measured transmission signal towards optical bistability without PI laser at the indicated Rabi frequency of $\Omega_c$. Arrows show the scan directions from red to blue detuning (red) and reverse (blue). (d) Color map of the transmission difference between two scan directions as a function of $\Omega_c$.}
    \label{Fig3}
\end{figure*}
Figure~\ref{Fig2} illustrates EIT spectra without (a) and with (b) PI laser by scanning the two-photon detuning $\Delta$. Importantly, a modulated EIT signal at $f\sim$11 kHz validates the construction of the self-induced Floquet system as we expected, $f \sim 1/T$. More details can be found in SM. Moreover, we extract the oscillation frequency under different static magnetic fields by locking the laser to a detuning of $-$8~MHz relative to the peak of each spectrum, as shown in Fig.~\ref{Fig2}(c). Each frequency point is obtained by a Fourier transform with a 16~ms time window of the oscillation signal. Inset demonstrates a 1~ms time dynamics of the probe transmission 
at $\textit{B}=14.88$~G. A solid line displays the linear fitting through the origin to the data. The linear scaling agrees well with the physical picture.

In order to probe the nonequilibrium dynamics in the parameter space, we first engineer the Rydberg excitation for different coupling Rabi frequency $\Omega_c$. Figure~\ref{Fig3}(a) displays three EIT spectra at an indicated $\Omega_c$, from which we can always find the oscillations at the frequency of $f$. 
However, above a certain value the sub-harmonic oscillation with frequency of $f$/2 appears, for example, at $\Omega_c/2\pi = $ 1.68~MHz, the middle curve of Fig.~\ref{Fig3}(a). By further increasing the Rabi frequency of $\Omega_c/2\pi$ to 2.11~MHz, the range of sub-harmonic oscillations extends, the up curve of Fig.~\ref{Fig3}(a). 
It is because when the coupling $\Omega_c$ is small, the system is in the weak interaction regime, where the dissipation dominates the dynamics of the system leading to a homogeneous phase. Then, the EIT spectrum displays oscillation with the internal frequency $f$ [see bottom curve of Fig.~\ref{Fig3}(a) with $\Omega_c/2\pi=1.08$~MHz], which is estimated by simulation of Fig.~\ref{Fig1}(d). However, as the coupling $\Omega_c$ increases, the excited Rydberg number increases and the system gets into the strong interaction bistable regime.  
Thus, the strong Rydberg atom interaction begins to compete with the internal driving by charges, and results in a subharmonic oscillation of the frequency $f/2$ [see the up curve of Fig.~\ref{Fig3}(a)] apart from the $f$ frequency component. This validates the theoretical simulation of Fig.~\ref{Fig1}(f), e.g. the DTC phase is established by the self-induced Floquet driving. 
To explore the phase transition, we do a series of measurements, as shown detailed spectra in Fig.~\ref{Fig3}(b), where $\Omega_c/2\pi$ is varied from 0.48~MHz to 2.42~MHz. Here, the white curve is a guideline to distinguish the $f$ component and the $f/$2 component.

As discussed in the theoretical model, there exists a link between the DTC phase and bistability. To reveal it from our experiment, we proceed to seek the bistable regime of the system without the PI laser beam. By scanning coupling laser frequency through the resonance Rydberg state from negative to positive detuning and vice versa. Figure~\ref{Fig3}(c) shows the stationary EIT spectra [other parameters are same as Fig.~\ref{Fig3}(a)]. 
At $\Omega_c/2\pi = $ 1.08~MHz, we observe almost overlapped EIT spectra. However, upon increasing $\Omega_c$, the EIT spectra become asymmetric, and optical bistability is observed. Moreover, the spectra become broadened when we further increase $\Omega_c$. With a series of measurements, the transmission difference between two opposite scan directions is sketched in Fig.~\ref{Fig3}(d). The measured results clearly show the phase transition from the homogeneous phase to the bistable phase. 
In combination with the information of Fig.~\ref{Fig3}(b) and (d), we conclude that bistability is required for generating $f/$2 component oscillations.

\begin{figure}[htbp]
\centering    
\includegraphics[width=\linewidth]{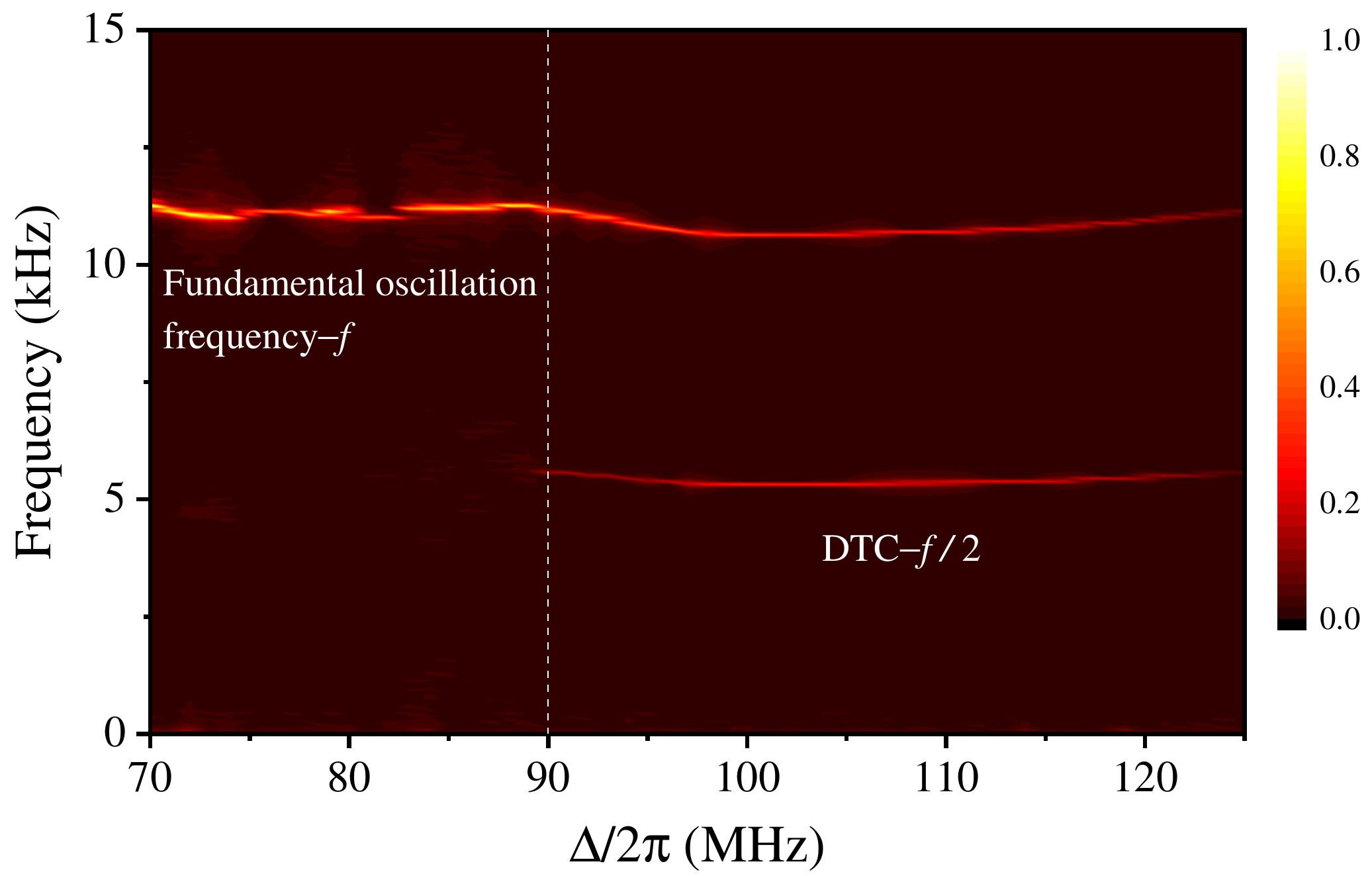}
\caption{Measured phase map of Fourier spectra as a function of the two-photon laser detuning $\Delta$ range from 70~MHz to 125~MHz at $\Omega_c = 2\pi \times$1.78~MHz. The critical point value is found with $\Delta\simeq90$~MHz (marked by the dashed line), indicating phase transition in such a system.} 
\label{Fig4}
\end{figure}

When fixing our system parameters in bistable regime where $\Omega_c = 2 \pi \times$1.78~MHz and $\delta_p$ = 2$\pi \times$110~MHz, we demonstrate that the phase diagram in terms of Fourier spectra of the persistent oscillation varies with two-photon detuning $\Delta$ [extracted from a time window of $16$~ms; see Fig.~\ref{Fig4}]. Here we lock the coupling laser to various laser detuning values via the offset locking technique. It is found that the frequency  $f$/2 component emerges when $\Delta$ exceeds the critical point value of about 90~MHz (marked by the dashed line), indicating the signature of nonequilibrium phase transition in this self-induced Floquet system. 

\textit{Conclusions}--We report the realization of a self-induced Floquet system in the thermal Rydberg gas. This is enabled by photoionizing the vapor gas in the presence of a static magnetic field, inducing periodic electric fields driving on the ground-Rydberg superpositions in a spatially separated channel. Based on such a platform, we investigate the nonequilibrium dynamics in the bistable regime. Essentially, we explore the link between the bistability and discrete time-crystalline phase on both the theoretical and experimental sides. Our work reveals the crucial role of the interplay of the strong interaction, dissipation and the associated multistability in the emergence of a discrete time crystalline. Our system emerges as a ubiquitous platform with a rich toolbox for exploring the exotic dynamic phases of many-body systems.

\textbf{Acknowledgements} 
This work was supported by the NSFC of China (grant no. U2341211, 12241408, 62175136, 12120101004, 12274131 and 12222409). H. S. acknowledges financial support from the Royal Society Newton International Fellowship Alumni follow-on funding (AL201024) of the United Kingdom. C.S.A. acknowledges the financial support provided by the UKRI, EPSRC grant reference number EP/V030280/1 (“Quantum optics using Rydberg polaritons”).

\bibliography{main}

\end{document}